\documentclass[aps,prl,floatfix,preprintnumbers,twocolumn,groupedaddress,nofootinbib]{revtex4-1}
\usepackage{amsmath,amsthm,amssymb,color,psfrag,url,latexsym,graphicx,epstopdf,slashed,xspace,hyperref,enumitem}
\usepackage{lineno}

\definecolor{darkred}{rgb}{0.6,0.0,0.0}
\definecolor{darkblue}{rgb}{0.0,0.0,0.5}
\definecolor{darkgreen}{rgb}{0.0,0.5,0.0}
\definecolor{brown}{rgb}{0.0,0.0,0.0}

\newcommand{\be}{\begin{equation}}
\newcommand{\ee}{\end{equation}}
\newcommand{\bea}{\begin{eqnarray}}
\newcommand{\eea}{\end{eqnarray}}

\begin{document}
\preprint{MIT-CTP 4952}
\title{Telescoping Jet Substructure}

\author{Yang-Ting Chien$^{a,d}$}
\email{ytchien@mit.edu\\}

\author{Alex Emerman$^{c,d,e}$}

\author{Shih-Chieh Hsu$^b$}

\author{Samuel Meehan$^b$}

\author{Zachary Montague$^{a,b}$}

\affiliation{
$^a$ Center for Theoretical Physics, Massachusetts Institute of Technology, Cambridge, MA 02139\\
$^b$ Department of Physics, University of Washington, Seattle, WA 98195\\
$^c$ Department of Physics, Columbia University, New York, NY 10027\\
$^d$ Theoretical Division, T-2, Los Alamos National Laboratory, Los Alamos, NM 87545\\
$^e$ Physics Department, Reed College, Portland, OR 97202
}

\date{\today}

\begin{abstract}
We introduce a novel jet substructure method which exploits the variation of observables with respect to a sampling of phase-space boundaries quantified by the variability. We apply this technique to identify boosted $W$ boson and top quark jets using telescoping subjets which utilizes information coming from subjet topology and that coming from subjet substructure. We find excellent performance of the variability, in particular its robustness against finite detector resolution. The extension to telescoping jet grooming and other telescoping jet substructure observables is also straightforward. This method provides a new direction in heavy particle tagging and suggests a systematic approach to the decomposition of jet substructure.
\end{abstract}
\maketitle

The Large Hadron Collider (LHC) has begun to probe physics above the electroweak scale, where the momenta of massive Standard Model particles are much larger than their invariant masses, resulting in hadronic decays of jets with prong-like substructures. Many jet substructure variables have been designed~\cite{Abdesselam:2010pt,Altheimer:2012mn,Altheimer:2013yza} and combined using multivariate techniques~\cite{Adams:2015hiv,Larkoski:2017jix,ATLAS-CONF-2017-064,Khachatryan:1955546} to identify such jets and increase the sensitivity to beyond the Standard Model physics. The ability to reconstruct the features of such jets accurately is obscured by the presence of additional proton-proton interactions, i.e. pileup, as well as the underlying event of the hard collision, both of which cause additional radiation to fall within the catchment area of the jet. Often, this radiation is removed through a grooming procedure, e.g. pruning~\cite{Ellis:2009su} or trimming~\cite{Krohn:2009th}. Jet substructure observables and grooming procedures target certain intuitive features of the radiation properties and often have tuneable parameters. For example, the pruning parameters $z_{\rm cut}$ and $D_{\rm cut}$ control the softness and noncollinearity of a discarded particle. Conventionally, one makes a single choice of parameters deemed optimal by some metric. However, such a choice may neglect the full information the entire observable class contains.
\newline \indent Recently, Q-jets \cite{Ellis:2012sn} introduced non-determinism in jet clustering. The procedure probes each jet multiple times and quantifies differences among pruned jets using the mass volatility. Later, telescoping jets \cite{Chien:2014hla} probed the radiation pattern surrounding the dominant energy flow with multiple angular resolutions $\{R_i\}$ and extracted the full information contained in jets at all angular scales. In this Letter, we apply telescoping jets to analyze a set of commonly used jet observables and grooming procedures. We demonstrate the feasibility of this method as applied to the identification of hadronically decaying $W$ bosons and top quarks, utilizing the variability of each observable induced by the variation of its parameters.
\newline \indent In hadronic boosted two-body resonance decays, such as that from a $W$ boson, the resonance mass $M$ introduces a two-prong structure in the jet at an angular scale $\Theta\approx 2M/p_T$ between the two prongs, where $p_T$ is the transverse momentum of the heavy particle. On the other hand, QCD jets initiated by isolated quarks and gluons do not have such a distinct scale. However, when examining jets with masses near $M\pm\Delta m$, QCD jets are also two-prong-like but with a more distended radiation pattern when $\Delta m\gg\Gamma$, where $\Gamma$ is the natural width of the resonance. Besides this nontrivial {\sl subjet topology}, the strong interaction dictates the formation of subjets with {\sl subjet substructures} and {\sl subjet superstructures} \cite{Gallicchio:2010sw} which are sensitive to the partonic origins of subjets.

In the case of boosted top quarks, the top mass ($M_t$) and the $W$ mass ($M_W$) are similar. Therefore $\Theta_t \approx 2M_t/p_T^t$ and $\Theta_W \approx 2M_W/p_T^W$ are comparable. This results in the generic three-prong structure in the hadronic top decay $t\rightarrow W+b \rightarrow q_1+q_2+b$. However, when examining jets with a mass near $M_t\pm \Delta m$ the selected QCD jets are, again, two-prong-like, so observables which distinguish three-prong jets from two-prong jets will help discriminate QCD jets from true top quark jets.

\begin{figure*}
    \includegraphics[width=2\columnwidth]{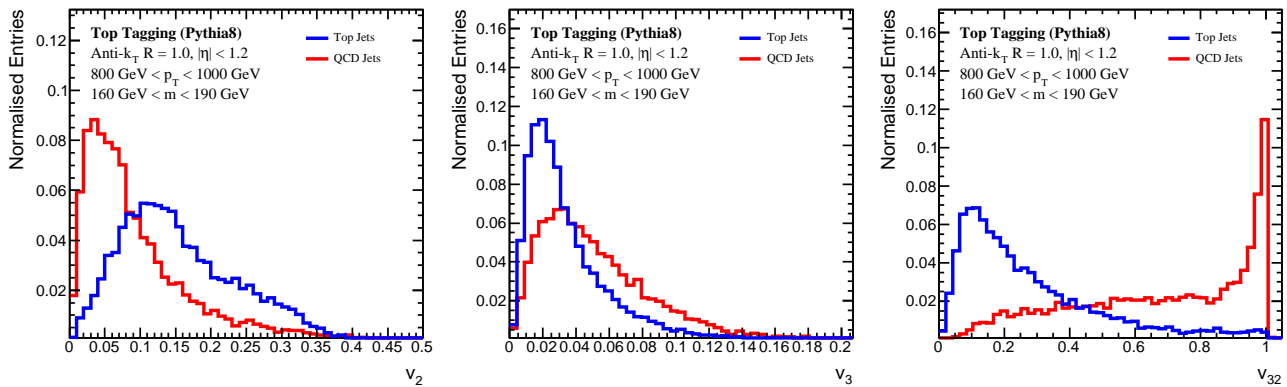}
    \caption{The distributions of the variabilities $v_2$ (left panel) and $v_3$ (middle panel), as well as their ratio $v_{32}$ (right panel) for top and QCD jets with $800~{\rm GeV} < p_T < 1~{\rm TeV}$ and $160~{\rm GeV} < m < 190~{\rm GeV}$ using the truth-particle information.}
\label{v42}
\end{figure*}

Given an arbitrary jet observable $\cal O$ with a parameter $a$,
the variation of the observable with respect to the sampling of parameters $\{a_i\}$ within $(a_{\rm min},a_{\rm max})$ is quantified by the coefficient of variation $v_{\cal O}^a$ defined as the ratio of the standard deviation and the mean of $\{{\cal O}_{a_i}\}$,
\be
    v_{\cal O}^a = \frac{\sigma({\cal O}_{a_i})}{\langle {\cal O}_{a_i}\rangle}\;.
\ee
The $v_{\cal O}^a$ observable is referred to as the {\sl variability}. Correlations among variations with respect to multiple varied parameters can also be explored. Analogous to the first derivative in calculus, the variability $v_{\cal O}^a$ measures the change of the observable $\cal O$ with respect to the change of the phase-space boundary set by the parameter $a$. Instead of combining observables with different parameters in a multivariate analysis, the variability can give a trend of the observable variation which itself can be used as a distinguishing feature to classify jets.

We focus on the variability of the jet mass with respect to varying the parameters which determine the jet constituents contributing to the jet mass. The sampling of the telescoping parameters is chosen to be uniform within the range $(a_{\rm min},a_{\rm max})$. We outline the procedure of \textit{telescoping subjets}: $N$ subjets are reconstructed exclusively around dominant energy flows within a jet. A similar method using the leading subjets in a reclustered jet was explored in \cite{Cui:2010km}. We groom the jets using the trimming algorithm with $R_{\rm sub}=0.3$ and $f_{\rm cut}=0.05$ to remove underlying event contaminations. Although we do not include pileup in our studies, using groomed jet constituents can also mitigate the pileup effect. We choose the subjet axes as the $N$-subjettiness axes \cite{Thaler:2011gf} with $\beta = 1$ and build subjets around them with radius $R_T$ within the voronoi regions \cite{Stewart:2010tn,Chien:2013kca,Stewart:2015waa,Thaler:2015xaa}. Particles are assigned to the nearest axis according to the distance $\Delta R_{ij}$ between the axis $\hat n_i$ and particle $p_j$,
\begin{equation}
    {\rm subjet}_{i} = \{p_j~|~\Delta R_{ij}<R_T~{\rm and}~\Delta R_{ij}<\Delta R_{kj}\;,~\forall k\neq i\},
\end{equation}
where $k$ is the index of the other axes $\hat n_k$. The variability $v_N$ of the invariant masses of the sum of $N$ subjets is reconstructed with the telescoping parameter $a = R_{T}\in (0.1, 1.0)\times R$. Note that $a_{\rm max}$ is chosen to be the jet radius $R$ to scan through the entire catchment area of the jet. On the other extreme, the dominant energy features will be lost if $a$ is too small, so $a_{\rm min}$ is chosen as $0.1\times R$. We focus on $N = $ 2 and 3 in $W$ and $N = $ 2, 3, and 4 in top tagging, but $N$ could be extended further for more exotic boosted topologies.

The generality of the telescoping algorithm allows a variety of other telescoping applications which are, however, beyond the scope of this Letter. For example, in \textit{telescoping pruning} one can fix $z_{\rm cut}$ and construct $v_{\rm prun}$, the variability of the pruned jet mass with the telescoping parameter $a$ in $D_{\rm cut} = a~ 2m_{\rm jet}/p_{T_{\rm jet}}$. In
\textit{telescoping trimming}, one can fix the subjet radius $R_{\rm sub}$ and construct $v_{\rm trim}$, the variability of the trimmed jet mass with the telescoping parameter $a = f_{\rm cut}$. One can also construct $v_{\tau_N}$, the variability of the $N$-subjettiness with the telescoping parameter $a = \beta$ \cite{Thaler:2010tr}.

Besides variabilities, useful angular observables, which encode information about subjet topology, and mass observables, which reveal the presence of specific decay products, can be obtained seamlessly from the telescoping subjet algorithm. For instance, in $W$ tagging with $N=2$, the subjet topology is affected by the jet mass cut, but $W$ and QCD jets can still have significantly different distributions for the angle $\theta_2$ between the two dominant energy flows. For top tagging with $N=3$, we consider the minimal angle $\theta_{\rm min}$ among the subjet axes. For QCD jets typically with two prongs, two of the three axes tend to be close to each other therefore $\theta_{\rm min}$ is expected to be small. For top jets with three prongs, this angle is distributed away from zero. Also, we attempt to identify the $W$ inside the top jet \cite{Thaler:2008ju,Kaplan:2008ie} by considering \textbf{$m_{W2}$}, the invariant mass of two of the three exclusive voronoi regions closest to the $W$ mass, and the variability $v_{m_{W2}}$ of the di-subjet invariant masses within those two regions.

\begin{figure*}
    \includegraphics[width=2\columnwidth]{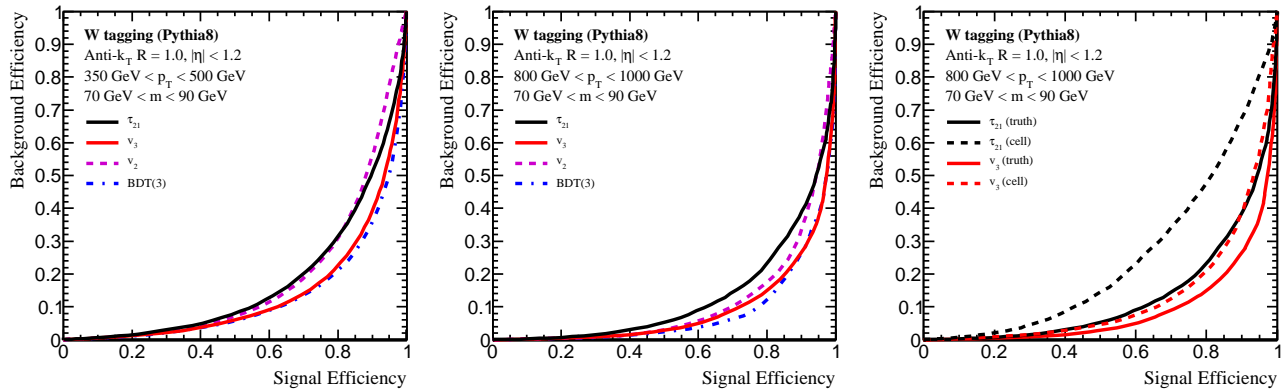}
    \caption{The $W$ tagging ROC curves of the variabilities $v_2$ and $v_3$;
    the BDT combinations of three telescoping subjets variables $\{v_2, v_3, \theta_2\}$; and the two-prong tagger $\tau_{21}=\tau_{2}/\tau_{1}$ in the $(300~{\rm GeV}, 500~{\rm GeV})$ jet $p_T$ bin (left panel) and the $(800~{\rm GeV}, 1~{\rm TeV})$ bin (middle panel). Right panel: ROC curves of $v_3$ and $\tau_{21}$ in the $(800~{\rm GeV}, 1~{\rm TeV})$ jet $p_T$ bin. Solid curves correspond to the ones with the truth-particle information, and the dashed curves are the ones using the pseudo-calorimeter cell particle information.}
\label{ROC_W}
\end{figure*}

The study is performed using samples generated from Monte Carlo simulations of proton-proton collisions at $\sqrt{s}=13$ TeV using \textsc{Pythia8} \cite{Sjostrand:2007gs}. Particles are clustered into jets with \textsc{FastJet} 3~\cite{Cacciari:2011ma} using the anti-$k_T$ algorithm \cite{Cacciari:2008gp} with $R=1.0$ and are required to be central with a pseudorapidity $|\eta|<1.2$. We consider two kinematic regimes where the jet $p_T$ is either between 350 GeV and 500 GeV or 800 GeV and 1 TeV. Signal $W$ boson and top quark jets are generated using decays of heavy Kaluza-Klein gravitons with invariant masses at 1 or 2 TeV for the two $p_T$ bins in fully hadronic $G^*\rightarrow W^+W^-$ and $G^*\rightarrow t\bar t$ processes. Background QCD jets are generated from the Standard Model dijet process. To study the impact of finite detector resolution, we compare the results with the particles clustered in pseudo-calorimeter $(\eta,\phi)$ cells of size $0.1\times 0.1$, with each cell momentum constructed with zero mass and direction from the primary vertex. A selection on the trimmed jet mass is made between 70 GeV and 90 GeV for $W$ tagging and between 160 GeV and 190 GeV for top tagging.

\begin{figure*}
    \includegraphics[width=2\columnwidth]{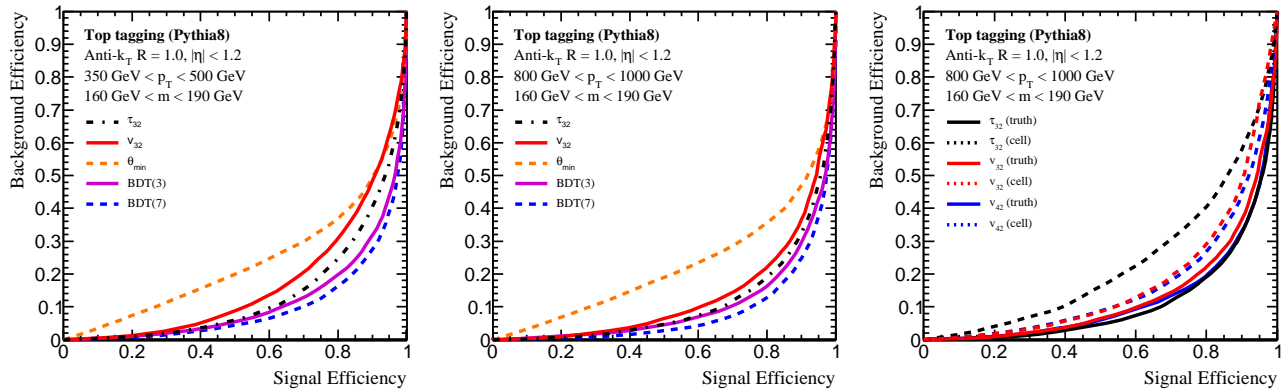}
    \caption{The top tagging ROC curves of the variability ratio $v_{32}$, the minimal angle among three subjets $\theta_{\rm min}$, the BDT combinations of three and seven telescoping subjets variables $\{m_{W2},v_2,v_3\}$ and $\{\theta_2,\theta_{\rm min},m_{W2},v_2,v_3,v_4,v_{m_{W2}}\}$, and the three-prong tagger $\tau_{32}=\tau_{3}/\tau_{2}$ in the $(300~{\rm GeV}, 500~{\rm GeV})$ jet $p_T$ bin (left panel) and the $(800~{\rm GeV}, 1~{\rm TeV})$ bin (middle panel). Right panel: ROC curves of $v_{32}$, $v_{42}$ and $\tau_{32}$ in the $(800~{\rm GeV}, 1~{\rm TeV})$ jet $p_T$ bin. Solid curves correspond to the ones with the truth-particle information, and the dashed curves are the ones using the pseudo-calorimeter cell particle information.}
\label{ROC_top}
\end{figure*}

To examine the complementarity of the information contained in the telescoping subjet variables, subsets of them are inputs for Boosted Decision Trees (BDTs) implemented in \textsc{TMVA} \cite{Hocker:2007ht}. For top tagging we also consider the ratio $v_{N2}$ between $v_N$ and $v_2$ for $N = $ 3, 4,
\be
    v_{N2}=\frac{v_N}{v_2}\;.
\ee
Shown in Figure~\ref{v42} are the distributions of $v_2$, $v_3$, and $v_{32}$ for top and QCD jets. We find that top jets have a broader $v_2$ distribution and a narrower $v_3$ distribution. The large variation of the jet mass when telescoping around the two subjet axes is caused by the transition of the $W$ from being partially reconstructed to fully reconstructed. There is not an intrinsic mass scale dictating the third hard emission for QCD jets. On the other hand, the three prongs inside top jets are quark-initiated subjets, whereas the subjets in QCD jets can have gluonic origins. Quark subjets are narrower than gluon subjets; therefore $v_3$ of top jets tend to be smaller. The $v_{32}$ observable has almost the same performance as the BDT with input $\{v_2,v_3\}$, suggesting that $v_{32}$ may be the optimal way of combining the two variabilities.

An interesting feature of $v_{32}$ is that it cuts off naturally at 1, most clearly seen in QCD jets. Crucially, $v_{3} \leq v_{2}$ in the collinear limit. The two-prong structure in QCD jets implies that $v_{2}$ and $v_{3}$ collect similar information. The third energy flow axis can not be displaced far from the two axes determined at $N=2$. Hence, little new information is collected by constructing a third subjet and the distribution of $v_{32}$ for QCD jets peaks at 1. In the case where there is a third, semi-hard emission, the emission is captured by all telescoping subjets at $N=3$ and does not induce the observable variation and so $v_{3} < v_{2}$. In general, for larger $N$, more particles are captured by-default and so the variability is expected to decrease ($v_{N+1}\leq v_{N}$).

The background efficiency as a function of the signal efficiency is illustrated by receiver operating characteristic (ROC) curves, where a lower curve indicates a better tagging performance. Shown in Figure~\ref{ROC_W} are the ROC curves of $v_2$, $v_3$, the BDT combinations of the telescoping subjet variables $\{v_2, v_3, \theta_2\}$, and the two-prong tagger $\tau_{21}=\tau_{2}/\tau_{1}$ in $W$ tagging. The left and middle panels correspond respectively to two jet $p_T$ regions of $(350~{\rm GeV}, 500~{\rm GeV})$ and $(800~{\rm GeV}, 1~{\rm TeV})$. Overall, the tagging performance increases at higher $p_T$, demonstrating the general advantage of applying telescoping jets to the boosted regime. In the right panel, we compare the tagging performance using truth particles and pseudo-calorimeter clusters, which degrade information about structures smaller than the cell size. We find excellent performance of $v_3$. Also, the $v_3$ observable is much more robust against this smearing, especially at high $p_T$, showing its qualitatively different feature compared to $\tau_{21}$. The $v_3$ observable utilizes the rapid depletion of radiation around the $W$ at larger angles in the boosted regime. This "$W$ isolation" effect is the manifestation of the fact that the $W$ carries zero color charge which affects the color structure of the subjets and the radiation pattern at large angles. The time dilation that occurs before $W$ hadronically decays can also result in a period of time in which no QCD radiation is emitted, while there is no such gap in the jet formation process for QCD jets. On the other hand, the fact that $v_3$ performs better than $v_2$ hints at the significance of a third, semi-hard emission in $W$ and QCD jets. The $v_3$ observable disentangles such emission when quantifying the isolation of $W$ jets.

Shown in Figure \ref{ROC_top} are the ROC curves for top tagging performance including $v_{N2}$ ($N = $ 3, 4), $\theta_{\rm min}$, the BDT combinations of telescoping subjets variables $\{v_2, v_3, m_{W2}\}$ and $\{\theta_2,\theta_{\rm min},m_{W2},v_2,v_3,v_4,v_{m_{W2}}\}$, and the three-prong tagger $\tau_{32}=\tau_{3}/\tau_{2}$. Again, the left and middle panels correspond to the two kinematic regimes $p_T \in (350~{\rm GeV}, 500~{\rm GeV})$ and $p_T \in (800~{\rm GeV}, 1~{\rm TeV})$, and we note tagging performance increases at higher $p_T$. In the right panel, the ROC curves plot both truth-particle and pseudo-calorimeter information. We find the excellent performance of $v_{N2}$ and its robustness against smearing, especially at high $p_T$ where the performance of the more conventionally used $\tau_{32}$ observable degrades dramatically. This indicates the qualitatively different features of $v_{N2}$ and a three-prong tagger. We also find the usefulness of including $m_{W2}$ in the minimal BDT combination which significantly increases the tagging performance. It is clear that the intrinsic mass scale $M_W$ within the top jet is a unique feature distinguishing itself from the QCD background. Similar to the fact that $v_3$ performs better than $v_2$ in $W$ tagging, the $v_{42}$ observable has a better performance than $v_{32}$, suggesting the significance of a fourth, semi-hard emission in top jets. One would also start to see the $W$ isolation within the top jet in the boosted regime.

To conclude, we introduce a qualitatively new jet substructure method using variability to quantify the change of observables with respect to a sampling of the phase-space boundaries in the observable definition. This technique is general and can be used to analyze arbitrary classes of jet substructure observables and grooming procedures. In this context of $W$ and top tagging, we find excellent performance of telescoping subjets quantified by $v_3$ in $W$ tagging and $v_{42}$ in top tagging. Furthermore, their robustness is found to be significantly better than more widely used $N$-prong taggers such as $N$-subjetiness via a comparison of the performance between reconstruction from using truth particles and from a pseudo-calorimeter.

The new physics messages we learn include the emergence of the isolation of $W$ jets at high $p_T$, which is a dominant feature over their two-prong structure. This is true for all other heavy, color-singlet Standard Model particles including the $Z$ and the Higgs boson. It would be promising to include such feature in their tagging strategies. The top jet also has features beyond the three-prong structure which can be exploited to increase tagging performance. The telescoping subjets provides a systematic framework within which one can construct qualitatively new jet substructure observables. This paves the road toward systematic jet studies using telescoping deconstruction \cite{Chien:2017decon}.

\section{Acknowledgements}
Y.-T. Chien would like to thank the organizers of the BOOST2015 conference where telescoping jet substructure was first presented. The authors thank Steve Ellis, Matthew Schwartz, Jesse Thaler and Nhan Tran for useful comments and suggestions. Y.-T. Chien was supported by the US Department of Energy (DOE), Office of Science under Contract No. DE-AC52-06NA25396, the DOE Early Career Program and the LHC Theory Initiative Postdoctoral Fellowship under the National Science Foundation grant PHY-1419008. A. Emerman was supported by the National Science Foundation under Grant No. PHY-1707971. S.-C. Hsu and S. Meehan were supported by the DOE Office of Science, Office of High Energy Physics Early Career Research program under Award Number DE-SC0015971. Z. Montague was supported by the University of Washington's Ernest M. Henley \text{\&} Elaine D. Henley Endowed Fellowship.

\bibliography{TJet_ref}
\end{document}